\documentclass[aps,prb,%
twocolumn,a4paper,showpacs,draft%
]{revtex4}

\usepackage[final]{graphicx}

\begin{document}

\title{Many-body Diagrammatic Expansion for the Exchange-Correlation Kernel 
in Time Dependent Density Functional Theory}
\author{I. V. Tokatly}
\email[e-mail: ]{ilya.tokatly@physik.uni-erlangen.de}
\altaffiliation{on leave from Moscow Institute of Electronic 
Technology, Zelenograd, 103498, Russia;}

\author{R. Stubner}
\author{O. Pankratov}

\affiliation{Lehrstuhl f\"ur Theoretische Festk\"orperphysik,
  Universit\"at Erlangen-N\"urnberg, Staudtstr.\ 7, 91058 Erlangen, Germany}

\date{August 1, 2001} 

\begin{abstract}
  A diagrammatic expansion for the dynamic exchange-correlation kernel
  $f_{xc}$ of time dependent density functional theory is formulated. It
  is shown that $f_{xc}$ has no singularities at Kohn-Sham transition
  energies in every order of the perturbation theory. However, it may
  diverge with the system size in extended systems. This signifies that
  any approximate perturbative substitute for $f_{xc}$ requires a
  consistent perturbative treatment of the response function to avoid
  uncontrollable errors in the many-body corrections to excitations
  energies.
\end{abstract}
 
\maketitle

The central problem of time dependent density functional theory (TDDFT)
\cite{RG} is to find an adequate approximation for the dynamic
exchange-correlation (xc) potential $v_{xc}$. In contrast to static DFT,
where the local density approximation (LDA) has been extremely
successful, no universal recipe for a dynamic $v_{xc}$ has been found,
and it remains unclear if such a recipe exists. The adiabatic LDA
(ALDA), which is most popular in practical TDDFT calculations, is valid,
by its nature, only in a quasistatic limit. This may suffice for a
real-time dynamics of melting \cite{melting} or desorption
\cite{desorp}, but is hardly relevant for electronic excitations in
insulators, where ALDA predicts the same erroneous band gaps as the
static LDA.\cite{sem} The improvement in excitation spectra of atoms and
molecules which has been obtained in TDDFT using ALDA or the optimized
effective potential (OEP) approximation is not indicative because in
small systems the correction to the Kohn-Sham excitation energies is
very small and approximations for $f_{xc}$ play a secondary role in
comparison to the accuracy of the static xc
potential.\cite{Petersilka2,mol}

In general, electron-hole ({e-h}) excitation energies can be found as
the poles of the density response function $\chi({\bf r},{\bf
  r'},\omega)$. TDDFT relates the exact $\chi(\omega)$ to the
susceptibility of non-interacting Kohn-Sham (KS) particles
$\chi_{S}(\omega)$ via the equation
\begin{equation}
\chi(\omega) = \chi_{S}(\omega) + \chi_{S}(\omega)
{\tilde V}(\omega)\chi(\omega)  
\label{1}
\end{equation}
with $\tilde V(\omega)=V_{C}+f_{xc}(\omega)$, where $V_{C}$ is the
Coulomb potential and the xc kernel $f_{xc}=\delta v_{xc}({\bf
  r},t)/\delta n({\bf r'},t')$ describes xc effects at the linear
response level.\cite{Petersilka1} Equation (\ref{1}), which
simultaneously accounts for the self-energy and the electron-hole
correlations, may seem like an attractive alternative to the very laborious
two-step approach involving a $GW$ calculation for one-particle states
with a subsequent solution of the Bethe-Salpeter (BS) equation for an
e-h pair.  Unfortunately, no approximation for $f_{xc}$ is available
which would be as efficient as LDA in the static DFT. In ALDA a
time-dependent density is simply inserted in the LDA xc potential, and
$f_{xc}$ becomes an instantaneous point interaction, which is
qualitatively different from a nonlocal and retarded xc kernel in
systems with an energy gap. In fact, it is {\it a priori} not clear
whether $f_{xc}$ allows any reasonable approximation, since its
analytical properties in a non-homogeneous system are not known.

From the viewpoint of the standard many-body formalism $\tilde
V(\omega)$ in Eq. (\ref{1}) acts as a mass operator for the density
propagator $\chi$, similar to the self-energy $\Sigma =
G_{0}^{-1}-G^{-1}$, with $G$ and $G_{0}$ being interacting and
non-interacting one-particle Green's functions. In contrast to the
Green's function, $\Sigma$ has no free-particle singularities in any
order of the perturbation theory, as it contains only irreducible
diagrams. Hence it allows perturbative approximations. Similarly, the
introduction of the ``density mass operator'' $\tilde V = \chi_{S}^{-1}
- \chi^{-1}$ is motivated only if the latter does not possess the e-h
singularities, which are present in $\chi_{S}$ and any finite order
approximation to $\chi$.

In this paper we use the KS-based many-body diagrammatic technique of
Ref.~\onlinecite{Tok2001} to derive a perturbative expansion for
$f_{xc}$. We find that the kernel is indeed regular at KS frequencies.
Yet, caution must be exercised in applying perturbative approximations
[such as OEP (Refs.~\onlinecite{Petersilka1,OEP})] for $f_{xc}$ in large systems to
avoid uncontrollable errors or even unphysical divergences.

We start with the equation which relates $\chi(\omega)$ to the proper
polarizability $\tilde \chi$
\begin{equation}
\chi(\omega) = \tilde \chi(\omega) + \tilde \chi(\omega)V_{C}\chi(\omega).  
\label{2}
\end{equation}
It is convenient to split $\tilde \chi$ as $\tilde
\chi=\chi_{S}+\pi_{xc}$ where the xc part $\pi_{xc}$ can be represented
as a series of graphs.\cite{Tok2001} The first-order contribution
$\pi_{xc}^{(1)}$ and examples of the second-order corrections
$\pi_{xc}^{(2)}$ are shown in Fig. 1, where solid and dashed lines stand
for the KS Green's functions and the Coulomb interaction respectively.
To the wiggled line is assigned the inverse KS susceptibility
$\chi_{S}^{-1}$. It describes the scattering of KS particles by the xc
potential $v_{xc}$. By the definition of $v_{xc}$, these processes
exactly compensate the change of the density due to the self energy
insertions in every order of the perturbation theory.\cite{Tok2001}

\begin{figure}
  \includegraphics[width=0.45\textwidth]{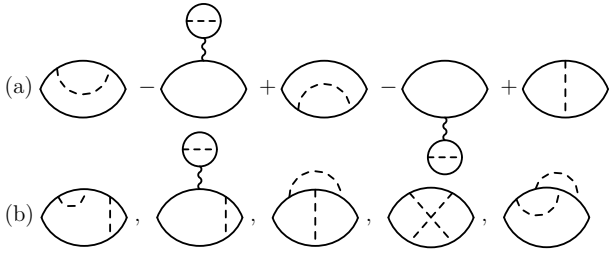}
  \caption{(a) The first-order correction $\pi_{xc}^{(1)}$ to the
    proper polarizability; (b) examples of the second-order corrections}
  \label{fig1}
\end{figure}

The graphical expansion of $f_{xc}$ can be obtained from the relation
$f_{xc}=\chi_{S}^{-1}-\tilde \chi^{-1}$ (Ref.~\onlinecite{Petersilka1}), which follows
from Eqs. (\ref{1}) and (\ref{2}). The formal expansion of this equation
in terms of $\pi_{xc}$ gives the series
\begin{equation}
  \label{3}
  f_{xc} = \chi_{S}^{-1}\pi_{xc}\chi_{S}^{-1} 
         - \chi_{S}^{-1}\pi_{xc}\chi_{S}^{-1}\pi_{xc}\chi_{S}^{-1} \dots 
\end{equation}
Inserting $\pi_{xc} = \pi_{xc}^{(1)} + \pi_{xc}^{(2)} + \pi_{xc}^{(3)} +
\dots$ and collecting the diagrams of the same order we obtain
$f_{xc}^{(1)}$, $f_{xc}^{(2)}$, etc. The $n$-th order contribution
$f_{xc}^{(n)}$ is a sum of the polarizability of the $n$-th order
$\pi_{xc}^{(n)}$ with two attached wiggled lines and all diagrams with
lower-order polarization loops connected by wiggled lines (Fig. 2).  A
closer inspection shows that all graphs with {\it internal} wiggled
lines can be generated from a limited set of graphs with no internal but
two {\it external} wiggled lines. This leads to the following rules for
constructing $f_{xc}^{(n)}$: (i) Draw all standard diagrams
\cite{LandauIX} for the proper polarization operator of the $n$-th order
and attach wiggled lines to external points (construction of parent
graphs). (ii) If possible, separate any given graph into two by cutting
two fermionic lines. Join the external fermionic lines of these parts,
connect them by a wiggled line and change the sign. Do not cut lines
attached to the same wiggled line. (iii) Apply (ii) to all possible
cuttings in all graphs, including those obtained previously. (iv) Keep
only nonequivalent graphs.  The arrows in Fig.~2 indicate application of
these rules to the second-order parent graph in the upper left
corner.\footnote{Interestingly, the rules (ii)-(iv) coincide with the
  diagrammatic rules for the xc potential $v_{xc}$
  (Ref.~\onlinecite{Tok2001}). The only difference is the choice of the
  parent graphs in (i) (loops with one or two external points for
  $v_{xc}$ and $f_{xc}$ respectively).  This by far nontrivial fact
  leads to the conjecture that these rules should hold for any
  functional derivative $\delta^{m}E_{xc}/\delta n^{m}$ of the xc energy
  $E_{xc}$.}

\begin{figure}[b]
  \includegraphics[width=0.45\textwidth]{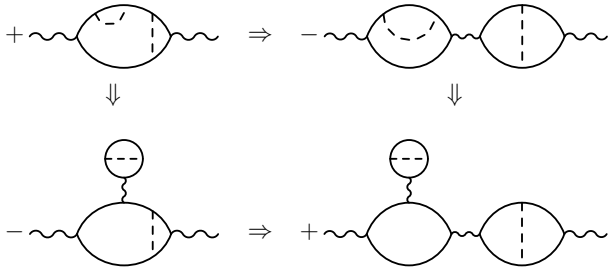}
  \caption{Examples of the second-order graphs for $f_{xc}$. The arrows
    indicate an application of the rules (ii)-(iv) to the 
    parent graph displayed in the upper left corner.}
  \label{fig2}
\end{figure}

The same rules apply to the perturbation series for the ``mass
operator'' $\tilde V$, except that in (i) the parent graphs are the
diagrams for the total response function $\chi$. As discussed above, the
introduction of $\tilde V$ (or $f_{xc}$) is justified only if this
function is free of singularities related to the KS e-h pairs in every
order of the perturbation theory. The analogy with the self-energy
$\Sigma$ does not, by itself, ensure this property, and we apply our
graphical method to prove that this is indeed the case.

Since the rules (ii)-(iv) deal only with the two-particle reducible
graphs, a partial summation of the diagrams with the help of one- and
two-particle irreducible elements (self-energies and vertices) is
possible. An example of a summation of all parent graphs with $l$ vertex
insertions (which divide each graph in $l+1$ blocks) and with $m_{k}$
($k=1,..,l+1$) self-energies in every block is depicted in Fig.~3a. It
is important that the diagrams generated by cutting two fermionic lines
with the same frequency (e.g. in Fig.~1 and Fig.~2) have a similar
structure. They all describe scattering of KS particles by the xc
potential. The sum of these graphs and the parent graphs is again a
diagram of the same type as in Fig.~3a with the self-energy
$\Sigma_{S}=G_{S}^{-1}-G^{-1}$. This definition of $\Sigma_{S}$ ensures
the equivalence of the KS and exact densities. The vertex $\Gamma$ is
defined in the usual way as the sum of the four-point functions which
are irreducible in the e-h channel.  After summation, the graph in
Fig.~3a can be considered as a new parent graph which generates further
diagrams for $\tilde V$ via cutting only parallel e-h lines.

Let us now consider the general parent graph of Fig.~3a at KS frequency
$\omega_{ij}=E_{i}-E_{j}$, where $E_{j}$ are the KS one-particle
energies.  This graph represents a $L$-th order correction (i.e.
containing $L$ irreducible elements) to the response function $\chi$.
When the frequency $\omega$ approaches $\omega_{ij}$, the graph in
Fig.~3a diverges. Integrating over intermediate frequencies in every
internal block we find that the most divergent term behaves as
$1/(\omega - \omega_{ij})^{(L+1)}$. Application of our rules to the
graph Fig.~3a generates a full set of diagrams for $\tilde V(\omega)$
i.e. the initial graph with only two external wiggled lines and the
diagrams obtained by all possible cuttings of parallel e-h lines. All
diagrams in this set have poles of various orders. However, they can be
grouped in such a way that all singularities cancel. The general proof
is straightforward but lengthy and will be published elsewhere. Here we
show this cancellation for a graph with $L$ vertices, but with no
self-energy insertions.

\begin{figure}
  \includegraphics[width=0.45\textwidth]{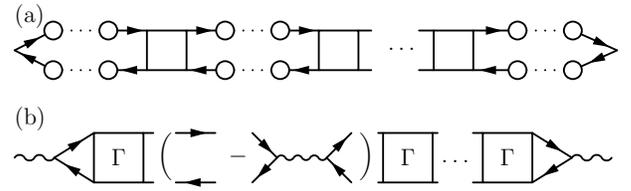}
  \caption{(a) General graph of the $L$-th order for the response
    functon $\chi$; (b) the diagram for the effective interaction which
    is produced from the general parent graph without self-energy insertions.}
  \label{fig3}
\end{figure}

Application of the rules (i)-(iv) to this graph leads to replacement of
all internal KS two-particle functions
$K_{S}(\omega,\varepsilon)=G_{S}(\omega+\varepsilon)G_{S}(\varepsilon)$
by
\begin{displaymath}
J(\omega,\varepsilon,\varepsilon') = K_{S}(\omega,\varepsilon)
\delta_{\varepsilon,\varepsilon'} -
K_{S}(\omega,\varepsilon)\chi_{S}^{-1}(\omega)K_{S}(\omega,\varepsilon'),
\end{displaymath}
where $\omega$ and $\varepsilon$ are transferred (external) and internal
frequencies respectively. The resulting contribution to $\tilde V$ is
shown graphically in Fig.~3b. The singularities at $\omega =
\omega_{ij}$ can potentially occur in the ``resonant'' part in every
internal block $J_{ij}(\omega,\varepsilon,\varepsilon')$, which contains
an electron in the state $i$ and a hole in the state $j$. A summation
over internal frequency gives for the ``dangerous'' contribution
\begin{equation}
\label{6}
\sum_{\varepsilon,\varepsilon'}J_{ij}(\omega,\varepsilon,\varepsilon') = 
\frac{|ij\rangle \langle ij|}{\omega  - \omega_{ij}} - 
\frac{|ij\rangle \{ ij|}{\omega  - \omega_{ij}}
\chi_{S}^{-1}(\omega)
\frac{|ij\} \langle ij|}{\omega - \omega_{ij}},
\end{equation}
where $|ij\rangle = \psi_{i}({\bf r})\psi_{j}^{*}({\bf r}')$ is the wave
function of the resonant e-h pair, $|ij\} = \psi_{i}({\bf
  r})\psi_{j}^{*}({\bf r})$ is the same function, but with equal
coordinates of the electron and the hole and $\psi_{i}({\bf r})$ are the
KS orbitals. Both terms in Eq.~(\ref{6}) are apparently singular. To
show that these divergences cancel, we single out the divergence in the
KS susceptibility $\chi_{S} = |ij\}\{ij|/(\omega - \omega_{ij}) +
\chi_{r}$, where $\chi_{r}$ is the regular part. For the inverse
$\chi_{S}^{-1}(\omega)$ we have
\begin{equation} 
\label{7}
\chi_{S}^{-1}(\omega) = \chi_{r}^{-1}(\omega) -
\frac{\chi_{r}^{-1}(\omega)|ij\}\{ij|\chi_{r}^{-1}(\omega)}
{\omega - \omega_{ij} + \{ij|\chi_{r}^{-1}(\omega)|ij\}}.
\end{equation}
Substitution of Eq.~(\ref{7}) to Eq.~(\ref{6}) gives a singularity-free result 
\begin{equation}
\label{8}
\sum_{\varepsilon}J_{ij}(\omega_{ij},\varepsilon) = 
\frac{|ij\rangle \langle ij|}{\{ij|\chi_{r}^{-1}(\omega_{ij})|ij\}}. 
\end{equation}
Similarly, zeroes of the external wiggled lines cancel the poles of the
end blocks in Fig.~3b.\footnote{To simplify the derivation we assumed
  that the resonant transition $\omega_{ij}$ is not degenerate. In the
  case of $M$-fold degeneracy, the denominator in Eq.~(\ref{8}) is
  replaced by the inverse of the $M\times M$ matrix
  $\{ij|\chi_{r}^{-1}|i'j'\}$, where $|ij\rangle$ and $|i'j'\rangle$
  belong to the set of degenerate states.} Thus the graph Fig.~3b is
regular at KS resonances in every order of the perturbation theory, and
$\tilde V(\omega)$ can be viewed as a mass operator similar to the
one-particle self energy $\Sigma$. There is, however, one important
difference. Whereas the self energy exactly reduces to the one-particle
irreducible elements, the effective interaction $\tilde V(\omega)$, even
after cancellation of singularities, still contains parts of the bare
two-particle propagator with the resonant denominator being replaced by
$\{ij|\chi_{r}^{-1}(\omega_{ij})|ij\}$ (see Eq.~(\ref{8})). Since
$\chi_{r}^{-1}({\bf r},{\bf r'})$ in general goes to zero at $|{\bf
  r}-{\bf r'}|\to \infty$ and the functions $|ij\}$ have a normalization
factor $\sim 1/V$, the matrix element $\{ij|\chi_{r}^{-1}|ij\}$ may
vanish with the increase of the system size. For example, in a 3D
semiconductor the inverse of the KS response function at $\omega=E_{g}$
has an asymptotic behavior $\chi_{r}^{-1}(E_{g},{\bf r},{\bf r'})\sim
1/|{\bf r}-{\bf r'}|$ (in this case $\chi_{r}$ means the principal value
of $\chi_{S}$). Therefore the matrix element
$\{cv|\chi_{r}^{-1}(E_{g})|cv\}$ vanishes as $V^{-1/3}$. This makes it
problematic to construct approximations for $f_{xc}$ using perturbation
theory, as is routinely done for $\Sigma$.

\begin{figure}
  \includegraphics[width=0.45\textwidth]{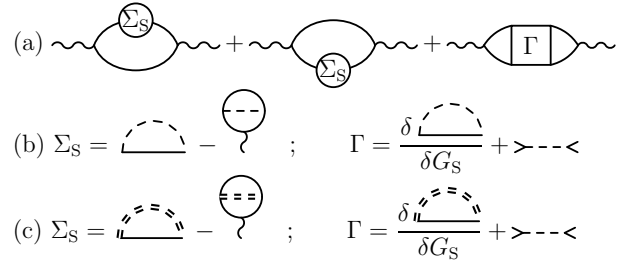}
  \caption{(a) Graphical representation of the general first-order
    approximation $\tilde V^{(1)}$ to the effective interaction; (b)
    self-energy $\Sigma_{S}$ and irreducible vertex $\Gamma$ for OEP; (c)
    $\Sigma_{S}$ and $\Gamma$ for RA approximation. The double dashed
    line stands for the screened Coulomb interaction.}
  \label{fig4}
\end{figure}

The simplest of the perturbative approximations corresponds to the first
order in irreducible elements $\Sigma_{S}$ and $\Gamma$ (Fig.~4a),
$\tilde V^{(1)}=\chi_{S}^{-1}\chi^{(1)}\chi_{S}^{-1}$, where
$\chi^{(1)}$ is the first-order polarization loop. This class of
approximations covers, for instance, the dynamic x-only OEP (Ref.~\onlinecite{OEP})
and the Richardson-Ashcroft approximation (RA).\cite{Rich-Ashcr} $\tilde
V_{OEP}(\omega)$ is given by the graph of Fig.~4a with $\Sigma_{S}$ and
$\Gamma$ taken in the first order in the Coulomb interaction $V_{C}$
(Ref.~\onlinecite{Tok2001}) in Fig.~4b. Similarly, $\tilde V_{RA}(\omega)$ has the
same form of Fig.~4a, but with the self-energy $\Sigma_{S}$ and the
vertex $\Gamma$ as shown in Fig.~4c. OEP and RA are state-of-the-art
approximations which accurately reproduce the correlation energy and
plasma excitations of a homogeneous electron gas.\cite{Ecor} However,
both approximations may give uncontrollable results for the e-h
excitation energies. In the following we show this by an explicit
calculation of the second-order correction to the KS excitation energy
$\omega_{ij}$.

The first-order (with respect to $\Sigma_{S}$ and $\Gamma$) correction
$\Delta \omega_{ij}^{(1)}= \omega_{ij}^{(1)}-\omega_{ij}$ is given by
\begin{equation}
  \label{9}
  \Delta\omega_{ij}^{(1)} = \{ij|\tilde{V}^{(1)}(\omega_{ij})|ij\}
                   \equiv \langle ij|\hat{W}|ij\rangle,
\end{equation}
where the operator $\hat{W}({\bf r}_{1},{\bf r}'_{1};{\bf r}_{2},{\bf
  r}'_{2})$ is defined as follows
\begin{eqnarray}
\nonumber
\hat{W} &=& \Sigma_{S}^{e}({\bf r}_{1},{\bf r}_{2}) 
             \delta({\bf r}'_{1}-{\bf r}'_{2}) - 
            \Sigma_{S}^{h}({\bf r}'_{1},{\bf r}'_{2})
            \delta({\bf r}_{1}-{\bf r}_{2})\\
&&{}+\Gamma^{eh}({\bf r}_{1},{\bf r}'_{1};{\bf r}_{2},{\bf r}'_{2}).
 \label{10}
\end{eqnarray}
The upper indexes in Eq.~(\ref{10}) indicate that the self-energies and
the vertex are taken on the e-h mass shell. The equations
(\ref{9})-(\ref{10}) give a generalization of the x-only
result,\cite{Tok2001} which is equivalent to the first-order of the
G\"orling-Levy perturbation theory.\cite{Gorling}

Similarly we obtain the second-order correction
\begin{eqnarray}
  & &  \Delta\omega_{ij}^{(2)} = \{ij|
    \tilde{V}^{(2)}(\omega_{ij}) |ij\} 
    + \Delta\omega_{ij}^{(1)} \{ij|
    \frac{\partial\tilde{V}^{(1)}}{\partial\omega}
    |ij\} \nonumber \\
  &+& \sum_{k,l\neq i,j} \frac{\{ij| \tilde{V}^{(1)}(\omega_{ij})
     |kl\}(n_{k}-n_{l}) \{kl|
    \tilde{V}^{(1)}(\omega_{ij}) |ij\}}{\omega_{ij} - \omega_{kl}},
  \label{11}
\end{eqnarray}
where $n_{k}$ is the occupation number of the $k$-th KS state and
$\tilde V^{(2)}(\omega)$ is the effective interaction of the second
order in $\Sigma_{S}$ and $\Gamma$. If we ignore $\tilde V^{(2)}$ and
iteratively solve Eq.~(\ref{1}) using only the first-order function
$\tilde V^{(1)}$ (e.g. in OEP or RA approximation) in place of the
kernel, the second-order correction Eq.~(\ref{11}) would contain only
the last two terms.  The last term can be written as
\begin{displaymath}
  \{ij| \tilde{V}^{(1)}(\omega_{ij}) \chi_{r}(\omega_{ij})
  \tilde{V}^{(1)}(\omega_{ij}) |ij\} =
  \frac{(\Delta\omega_{ij}^{(1)})^{2}}{\{ij| \chi_{r}^{-1}(\omega_{ij}) |ij\}},
\end{displaymath}
where we used Eq.~(\ref{9}) and Eq.~(\ref{7}). A similar calculation of
the second term in Eq.~(\ref{11}) shows that it contains the same factor
$\{ij|\chi_{r}^{-1} |ij\}^{-1}$. Hence both terms in Eq.~(\ref{11})
which originate from $\tilde V^{(1)}$, and which would be the only
contributions in OEP and RA, have exactly the same denominator as the
higher-order diagrams after the cancellation of the e-h singularities
(see Eq.~(\ref{8})). Clearly this denominator must appear in $\tilde
V^{(2)}$ as well. The calculation shows that the first term in
Eq.~(\ref{11}) indeed takes the form
\begin{eqnarray}
\nonumber
& &\{ij|\tilde V^{(2)}(\omega_{ij})|ij\} = \Delta_{ij}^{(2)} -
\Delta\omega_{ij}^{(1)} \{ij|
    \frac{\partial\tilde{V}^{(1)}}{\partial\omega}
    |ij\}\\
&&{}- \{ij| \tilde{V}^{(1)}(\omega_{ij}) \chi_{r}(\omega_{ij})
\tilde{V}^{(1)}(\omega_{ij}) |ij\},
\label{12}
\end{eqnarray}
where
\begin{equation}
\Delta_{ij}^{(2)} = \sum_{k,l\ne ij}\frac
{\langle ij|\hat{W}|kl\rangle\langle kl|\hat{W}|ij\rangle}
{\omega_{ij} - \omega_{kl}} + \Delta\omega_{ij}^{(1)} \{ij|
    \frac{\partial\hat{W}}{\partial\omega}
    |ij\}.
\label{13}
\end{equation} 
The second and the third terms in Eq.~(\ref{12}) cancel the second and
the third terms in Eq.~(\ref{11}) and the second-order energy shift is
simply given by Eq.~(\ref{13}). This also follows from a direct
perturbative solution of the BS equation. The cancellation found above
is not accidental and can be proven in every order of the perturbation
theory.

Thus any finite-order approximation for $\tilde V$ requires a consistent
perturbative treatment of Eq.~(\ref{1}) to obtain a meaningful energy
shift. In particular, the first-order approximations like OEP or RA are
appropriate for the calculation of the excitation energies only in the
first order in $\Sigma_{S}$ and $\Gamma$. The naive second-order
correction, (last two terms in Eq.~(\ref{11})) and higher-order
corrections are wrong and can even diverge in extended systems. A
summation of all orders (exact solution of Eq.~(\ref{1})) may produce a
finite result, but it will contain uncontrollable errors. In recent
TDDFT calculations for atoms it has been indeed observed that exact (in
contrast to perturbative) solution of Eq.~(\ref{1}) does not necessarily
improve the results.\cite{Petersilka2}

This work has been supported by the Deutsche Forschungsgemeinschaft
under Grant PA 516/2-1. O.~P. is grateful for a partial support from the
U.S. Department of Energy, Office of Basic Energy Sciences, Division of
Materials Science by the University of California Lawrence Livermore
National Laboratory under contract No. W-7405-Eng-48. The work of I.~T.
was partly supported by the Russian Federal Program ``Integration''.


\begin{thebibliography}[99]
\bibitem{RG}E.~Runge and E.~K.~U.~Gross, Phys. Rev. Lett. {\bf 52}, 997 (1984).
\bibitem{melting} J. Theilhaber, Phys. Rev. B {\bf 46}, 12990 (1992). 
\bibitem{desorp} Y. Miyamoto and O. Sugino, Phys. Rev. B {\bf 62},
  2039 (2000). 
\bibitem{sem}
F.~Kootstra, P.~L.~de~Boeij, and J.~G.~Snijders,
  J. Phys. Chem. {\bf 112}, 6517 (2000).
\bibitem{Petersilka2}M. Petersilka, E.~K.~U.~Gross, and K.~Burke, 
Int. J. Quantum Chem. {\bf 80}, 534 (2000).
\bibitem{mol}
M.~E.~Casida and D.~R.~Salahub, Journ. Chem. Phys. {\bf 113}, 8918 (2000). 
\bibitem{Petersilka1}M.~Petersilka, U. J. Gossmann and
  E.~K.~U.~Gross, Phys. Rev. Lett. {\bf 76}, 1212 (1996).
\bibitem{Tok2001}I.~V.~Tokatly and O.~Pankratov,
  Phys. Rev. Lett. {\bf 86}, 2078 (2001).
\bibitem{OEP} A.~G\"orling, Int. J. Quant. Chem. {\bf 69}, 265
    (1998).  
\bibitem{LandauIX} E.~M.~Lifshitz, L.~P.~Pitaewski, {\it Statistical
  Physics, Part 2}, Course of Theoretical Physics, Vol. 9 (Pergamon,
  New York, 1980).
\bibitem{Rich-Ashcr}C.~F.~Richardson and N.~W.~Ashcroft,
  Phys. Rev. B {\bf 50}, 8170 (1994). 
\bibitem{Ecor} M.~Lein, E.~K.~U.~Gross and J.~P.~Perdew,
  Phys. Rev. B {\bf 61}, 13431 (2001); 
  K.~Tatarczyk, A.~Schindlmayr and M.~Scheffler, Phys. Rev. B {\bf
    63}, 235106 (2001). 
\bibitem{Gorling}A. G\"orling, Phys. Rev. A {\bf 54}, 3912 (1996).

\end{thebibliography}
\end{document}